\documentclass[aps,amsfonts,prl,preprint,showpacs,nobibnotes,nofootinbib,tightenlines]{revtex4}

\usepackage{epsf}

\def\be{\begin{equation}}
\def\ee{\end{equation}}
\def\bea{\begin{eqnarray}}
\def\eea{\end{eqnarray}}

\def\sech{\mathop{\rm sech}\nolimits}
\def\Legendre{{\rm P}}

\begin{document}

\title{Static Negative Energies Near a Domain Wall}

\author{Ken D.~Olum}
\email{kdo@cosmos.phy.tufts.edu}
\affiliation{Department of Physics and Astronomy \\ Tufts
University \\ Medford, MA  02155}

\author{Noah Graham}
\email{graham@physics.ucla.edu}
\affiliation{Department of Physics and Astronomy \\
University of California at Los Angeles \\ Los Angeles, CA  90095}

\begin{abstract}
We show that a system of a domain wall coupled to a scalar field has
static negative energy density at certain distances from the domain
wall.  This system provides a simple, explicit example of violation of
the averaged weak energy condition and the quantum inequalities by
interacting quantum fields.  Unlike idealized systems with boundary
conditions or external background fields, this calculation is
implemented precisely in renormalized quantum field theory with the
energy necessary to support the background field included
self-consistently.
\end{abstract}

\preprint{gr-qc/0205134}
\preprint{UCLA/02/TEP/10}

\maketitle

\paragraph{Introduction}
In the absence of any restriction on the matter stress-energy tensor
$T^{\mu\nu}$, general relativity permits the construction of arbitrary
spacetime geometries.  One simply writes down the desired geometry and
solves Einstein's equation in reverse to determine $T^{\mu\nu}$.  Thus
it appears that the feasibility of producing exotic situations such as
closed timelike curves (however see
\cite{Krasnikov:2001ci,Krasnikov:2001cj}) and superluminal travel
depends on whether energy conditions restrict $T^{\mu\nu}$.
In particular, if one assumes the weak energy condition,
$T_{\mu\nu} V^\mu V^\nu\ge0$ for all timelike vectors $V^\mu$, 
or equivalently that no observer sees negative energy
density, then one can show that superluminal travel \cite{Olum:1998mu}
and the construction of closed timelike curves \cite{tip76,tip77,cpc}
are impossible.

The weak energy condition is obeyed by the usual classical
fields\footnote{It is, however, violated by non-minimally coupled
scalar fields \cite{Bekenstein:annals}.}, but quantum
fields can violate it.  Perhaps the simplest example of weak energy
condition violation is a superposition of the vacuum and a single mode
with 2 photons.  The negative energy densities are confined to
particular regions of space and particular periods of time.  In this system,
and in any system made from free fields \cite{Klinkhammer}, the energy
density satisfies the averaged weak energy condition,
\be\label{eqn:awec}
\int_{\cal G} T_{\mu\nu} V^\mu V^\nu\ge0\,,
\ee
where the integral is taken over a complete timelike geodesic
${\cal G}$ with tangent vector $V^\mu$.  This energy density also
satisfies quantum inequalities \cite{F&Ro97} of the
form
\be
\int \rho ({\bf x}, t) W(t) dt\ge-ct_0^{-d}
\ee
where $\rho =T_{00}$, $W$ is a window function of width $t_0$, $d$ is
the number of spacetime dimensions, and $c$ is a constant depending on
$d$, the type of field being considered, and the particular shape of
$W$.

On the other hand, the best-known system exhibiting negative energy
density is the Casimir problem.  Casimir \cite{Casimir} computed the energy
density of the quantum electromagnetic vacuum between perfectly
conducting plates separated by a distance $d$ and found
\be
\rho = -\frac{\pi^2}{720d^4}\,.
\ee
Laboratory measurements \cite{Bressi:2002fr} of the force associated
with this energy have found good agreement with Casimir's result.
While a question has been raised \cite{Helfer:1998pi} whether the
energy density between metal plates (as opposed to idealized perfect
conductors) is actually negative, it does appear to be so
\cite{Sopova:2002cs} as long as the separation of the plates is large
enough.  Since the negative energy density in the Casimir effect is
static, it can be averaged for arbitrarily long times.  Thus the Casimir
effect violates the averaged weak energy condition and the quantum
inequalities.

One way to think of the Casimir effect is as the energy of the
electromagnetic vacuum with specified boundary conditions or with
interaction with fixed materials.  In that model, the electromagnetic
field energy is subject to ``difference quantum inequalities''
\cite{F&Ro95}, which restrict the energy density to be not much more
negative than that in the vacuum with the specified conditions.
However, one can also think of the Casimir effect as the energy of a
system of coupled fields, including both the electromagnetic field and
the fields of the matter that makes up the plates.  In that case, the
Casimir system is just a particular excitation of some interacting
quantum fields above the vacuum.  It contains static negative energy
densities and is not restricted by any quantum inequality, because
those apply only to free fields.

Unfortunately, the actual Casimir system is quite complicated, and to
be certain to understand it one must take into account many effects
associated with physical metals, such as the true dispersion
relation and surface roughness.  This letter demonstrates the same
phenomenon of static negative energy density in a simpler system,
consisting only of two scalar fields in 2+1 dimensions.  Negative
energies have appeared previously in calculations of quantum energy
densities (see for example \cite{Bordag:1996}); our emphasis here is that
the complete energy density, including the energy required to support
the background field, is negative in a self-consistent approximation
with definite renormalization conditions.

\paragraph{Model}
In order to have a system of scalar fields that is static and does not
dissipate, we will use a topological defect.  For simplicity of
calculation and similarity to the Casimir problem we will use a domain
wall, and to decrease the number of divergences requiring
renormalization we will work in 2+1 dimensions.  We thus define a real
scalar field $\chi$ to form the domain wall and a second real scalar
field $\phi$ whose interactions with $\chi$ will produce the negative
energy density.  The Lagrangian is
\be
{\cal L} = \frac{1}{2}\left[\partial_\mu\chi\partial^\mu\chi
+\partial_\mu\phi\partial^\mu\phi +U(\chi,\phi)\right]
\ee
with
\be
U(\chi,\phi) = \lambda (\chi^2-\eta^2)^2 +m^2\
(1-\chi^2/\eta^2)\phi^2+\beta\phi^4\,.
\ee

With $\beta >m^4/(4\lambda\eta^4)$ we find that $U$ is positive
definite, and the classical ground state is given by $\phi = 0$ and
$\chi = \eta$ or $\chi = -\eta$.  If we specify different vacua for
$x\to\infty$ and $x\to -\infty$, we find a static classical domain
wall solution.  Taking the wall to lie on the $y$ axis, we find
\be
\chi (x) =\eta\tanh (x/a)
\ee
where $a =1/(\sqrt{\lambda}\eta)$.
The wall is invariant under translations and boosts in the $y$ direction.
It has classical energy density
\be\label{eqn:classical}
\rho(x) = 2\lambda\eta^4\sech^4(x/a)\,.
\ee

Now we quantize our fields.  If we work in the regime where
$m\ll\sqrt{\lambda}\eta$, the back reaction due to the quantized
$\phi$ will have a negligible effect on the domain wall.  We can thus
consider the domain wall, for the purposes of the calculation, to provide
a fixed background potential for $\phi$.  The effective Lagrangian is then
\be
{\cal L}_\phi =\frac{1}{2}\left[\partial_\mu\phi\partial^\mu\phi -V
(x)\phi^2\right]
\ee
where the potential
\be\label{eqn:potential}
V (x) = m^2(1-\chi^2/\eta^2) = m^2\sech^2(x/a)
\ee
acts as a position-dependent mass term for $\phi$.  Quantizing $\chi$
produces a correction to the shape of the domain wall and to its
energy, but the energy density still falls exponentially as one goes
away from the center of the wall.  The change of shape will affect the
Casimir energy associated with $\phi$, but only at higher order, which
we will not consider here \cite{Bashinsky}.

\paragraph{A simple argument}
We would like to show that a negative energy density exists
somewhere.  The energy density in the background potential can be
calculated exactly, and we will do so below, but a detailed calculation
is not necessary to demonstrate the existence of negative energy
densities.

Suppose that instead of the background potential we had a perfectly
reflecting boundary at $x = 0$, i.e., $\phi(0) = 0$.  Then the
computation would be straightforward and the energy density at $x$
would be
\be\label{eqn:mirror}
\rho_\phi(x) =-\frac{1}{32\pi x^3}
\ee
In this computation the main contribution to the energy density at
$x$ comes from those modes whose wavelength $\lambda$ is similar to
$x$.  If we take $x$ sufficiently large, then we will be interested in
large $\lambda$, and for sufficiently large $\lambda$, any potential
barrier is perfectly reflecting\footnote{The only exceptions to this
rule are potentials with a bound state precisely at threshold 
\cite{Barton,Graham99}, which include
reflectionless potentials.  We will not consider this exceptional case.}
Therefore, for sufficiently large $x$, the energy in
$\phi$ approaches the form of Eq.\ (\ref{eqn:mirror}).  To this energy
we must add the energy of the $\chi$ field, given in the classical
approximation by Eq.\ (\ref{eqn:classical}), from which
\be
\rho_\chi(x)\sim e^{-4 |x|/a}\,.
\ee
Even if we take into account quantum corrections to the domain wall
profile, we still expect $\rho_\chi$ to decline exponentially away
from the wall because $\chi$ is a massive field.

Thus the positive energy density associated with the wall declines
exponentially, while the negative energy density associated with
$\phi$ declines only as a power law.  For large enough $x$, the
negative energy will dominate, and the total energy will approach the
form of Eq.\ (\ref{eqn:mirror}).

\paragraph{Calculation}
The general calculation of the Casimir energy density for a scalar
field with a background potential will be presented elsewhere
\cite{Graham:coming}.  The energy density can be computed from the Green's
function for the given potential in scattering theory,
\be\label{eqn:energygreen}
\rho(x) =
-{1\over 8\pi}\int_0^\infty d\kappa\left[
2\kappa^3G(x, x, i\kappa)-\kappa^2+\frac{V (x)}{2}
-\kappa\frac{d^2}{dx^2} G(x, x, i\kappa)\right]
\ee
where $G (x, x', k)$ is the Green's function that satisfies
\be
-G''(x,x',k) +V(x)G(x,x',k) - k^2G(x,x',k) =\delta (x-x')
\ee
and has only outgoing waves ($\sim e^{ik|x|}$) at infinity.

\begin{figure}
\begin{center}
\leavevmode\epsfbox{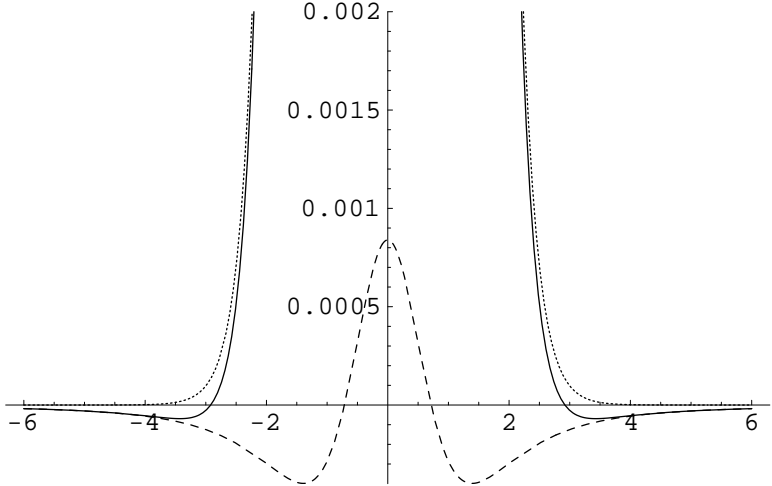}
\end{center}
\caption{Energy density in the wall (dotted) and the field $\phi$
(dashed) and the total energy density (solid) in units where $\eta =
1$ for parameters $\lambda = 1$ and $m^2 = 0.1$.  For sufficiently
large values of $x$, the total energy density is negative.}
\label{fig:wall}
\end{figure}

The problem of the potential of Eq.\ (\ref{eqn:potential}) can be
solved exactly.  The normal modes are associated Legendre functions
and the Green's function is
\be
G(x,x',k) = \frac{a}{2}\Gamma (1+s-ika)\Gamma (-s-ika)
\Legendre^{ika}_s (\tanh (x_>/a)) \Legendre^{ika}_s (-\tanh (x_</a))
\ee
where $x_<$ and $x_>$ are respectively the smaller and the larger of
$x$ and $x '$, $\Legendre^\mu_\nu (x)$ is the associated Legendre
function defined as in \cite{Bateman:v1} for $-1 < x < 1$, and $s =
(\sqrt{1-4m^2a^2} -1)/2$.
We have
\be\label{eqn:Green}
G(x,x,i\kappa) 
=\frac{a}{2}\Gamma (1+s+\kappa a)\Gamma (-s+\kappa a)
\Legendre^{-\kappa a}_s (\tanh (x/a)) \Legendre^{-\kappa a}_s (- \tanh (x/a))\,.
\ee
If we put Eq.\ (\ref{eqn:Green}) into Eq.\ (\ref{eqn:energygreen}) and
introduce the dimensionless variables $q =\kappa a$ and $y = x/a$ and the
parameter $v =m^2a^2$, we get
\bea
-\frac{1}{8\pi a^3}\int_0^\infty dq\bigg[& &
\Gamma (1+s+q)\Gamma (-s+q)
\left(q^3-\frac{q}{2}\frac{d^2}{dy^2}\right)
\left(\Legendre^{-q}_s (\tanh y) \Legendre^{-q}_s (-\tanh y)\right)\\
& &-q^2+{v\over 2}\sech^2y\bigg]
\eea
We are concerned with the regime where $m$ is small compared to the
inverse width of the wall, $1/a$, so $v\ll1$.
Fig.\ \ref{fig:wall} shows the energy density in the case where $v =
0.1$.  For $y\agt3$, the energy density is negative, as predicted
above.

\paragraph{Discussion}

We have shown a specific example of two interacting scalar fields
whose energy density is static and negative in certain regions.  Since
the system is static, one can average over as much time as one
chooses, and thus the system violates the averaged weak energy
condition and the quantum inequalities.  From this system (as from the
Casimir effect with physical plates)  one sees that the averaged weak
energy condition and the quantum inequalities are simply not correct
in the case of interacting fields, so the failure of attempts to prove
them is not due merely to technical reasons.

The present system does, however, satisfy the averaged null energy
condition, given by Eq.\ (\ref{eqn:awec}) with $V^\mu$ null, which is
sufficient to rule out superluminal travel and the construction of
time machines.  It is obeyed because if the geodesic runs parallel to
the domain wall, then the positive pressure cancels the negative
energy density and $T_{\mu\nu} V^\mu V^\nu = 0$, while if the geodesic
is not parallel to the wall it must cross through the region of high
positive energy.  It is not clear at this point whether some similar
arrangement, such as a domain wall in 3+1 dimensions with a hole,
might violate the averaged null energy condition for certain geodesics.

\paragraph{Acknowledgments}

We would like to thank J. J. Blanco-Pillado and Ruben Cordero for
assistance and Larry Ford, Bob Jaffe, Vishesh Khemani, Markus Quandt,
Tom Roman, Marco Scandurra, Xavier Siemens, Alexander Vilenkin, and Herbert
Weigel for helpful conversations.  N.~G. is supported by the
U.S.~Department of Energy (D.O.E.) under cooperative research
agreement \#DE-FG03-91ER40662.  K.~D.~O. is supported in part by the
National Science Foundation.

\bibliographystyle{apsrev}
\bibliography{gr}

\end{document}